\documentclass[conference]{IEEEtran}
\IEEEoverridecommandlockouts
\usepackage{amsmath,amssymb,amsfonts}
\usepackage{algorithmic}
\usepackage{graphicx}
\usepackage{textcomp}
\usepackage{xcolor}
\usepackage{ulem}
\usepackage[varg]{newtxmath}
\usepackage[switch]{lineno}
\usepackage{bm}
\usepackage{graphicx}
\usepackage{booktabs}
\usepackage{subfigure}
\usepackage{etoolbox}
\usepackage{caption}
\usepackage{mathtools}
\usepackage{cite}
\usepackage{color}
\usepackage{cancel}
\usepackage{verbatim}
\newtheorem{definition}{Definition}
\usepackage{epstopdf}
\usepackage{color}
\usepackage{amsmath}
\usepackage{array,graphicx,subfigure}
\usepackage{multirow}

\newcommand{\cut}[1]{}

\newcommand{\etal}{\textit{et al}.}

\def\BibTeX{{\rm B\kern-.05em{\sc i\kern-.025em b}\kern-.08em
    T\kern-.1667em\lower.7ex\hbox{E}\kern-.125emX}}

\begin{document}

\title{Finding Salient Context based on Semantic Matching for Relevance Ranking}

\author{\IEEEauthorblockN{1\textsuperscript{st} Yuanyuan Qi}
\IEEEauthorblockA{\textit{Beijing University of Posts and Telecommunications}\\
qiyuanyuan@bupt.edu.cn}
\and
\IEEEauthorblockN{2\textsuperscript{nd} Jiayue Zhang}
\IEEEauthorblockA{\textit{Beijing University of Technology}\\
zhangjiayue@bjut.edu.cn}
\and
\IEEEauthorblockN{3\textsuperscript{rd} Weiran Xu}
\IEEEauthorblockA{\textit{Beijing University of Posts and Telecommunications}\\
xuweiran@bupt.edu.cn}
\and
\IEEEauthorblockN{4\textsuperscript{th} Jun Guo}
\IEEEauthorblockA{\textit{Beijing University of Posts and Telecommunications}\\
Jun Guo}
}
\maketitle

\begin{abstract}
In this paper, we propose a salient-context based semantic matching method to improve relevance ranking in information retrieval. We first propose a new notion of salient context and then define how to measure it. Then we show how the most salient context can be located with a sliding window technique. Finally, we use the semantic similarity between a query term and the most salient context terms in a corpus of documents to rank those documents. Experiments on various collections from TREC show the effectiveness of our model compared to the state-of-the-art methods.
\end{abstract}

\begin{IEEEkeywords}
keywords matching, contextual salience, semantic matching.
\end{IEEEkeywords}

\section{Introduction}
As the core of understanding multimedia, semantic matching plays the role of bridge to connect different forms of content, such as text, image, video and audio, etc. Before semantic matching came into existence, the conventional keywords matching methods have been dominant for a long time, says, in Information Retrieval (IR)~\cite{manning2010introduction}. They fail, however, to capture the query term's fine-grained contextual information. The missing contextual information results in the term-mismatching problem due to the word ambiguity issue. To deal with this problem, varieties of neural IR models, which are often called semantic matching, have been proposed to incorporate context information by embedded representation ~\cite{NIR 2018}. Some methods consider the whole document as a global context and embed it into one vector. The query term is  embedded into a similar vector, and these vectors are  used to calculate the relevance between term and document~\cite{CDSSM}. Other methods consider a certain scope around the keyword as the local context. Only this local context is encoded into embedding vectors and used to compute the relevance~\cite{hui2018co}. Both parties have made important efforts to do semantic matching, but we believe that the retrieved documents can fit the query terms even better. The global context methods fail to capture the individual interactions between the query and the document terms since the whole document is  encoded into one vector. The latter group does  not have this problem, but it still leaves the mismatching problem unsolved.

\begin{figure}
\includegraphics[scale=0.35]{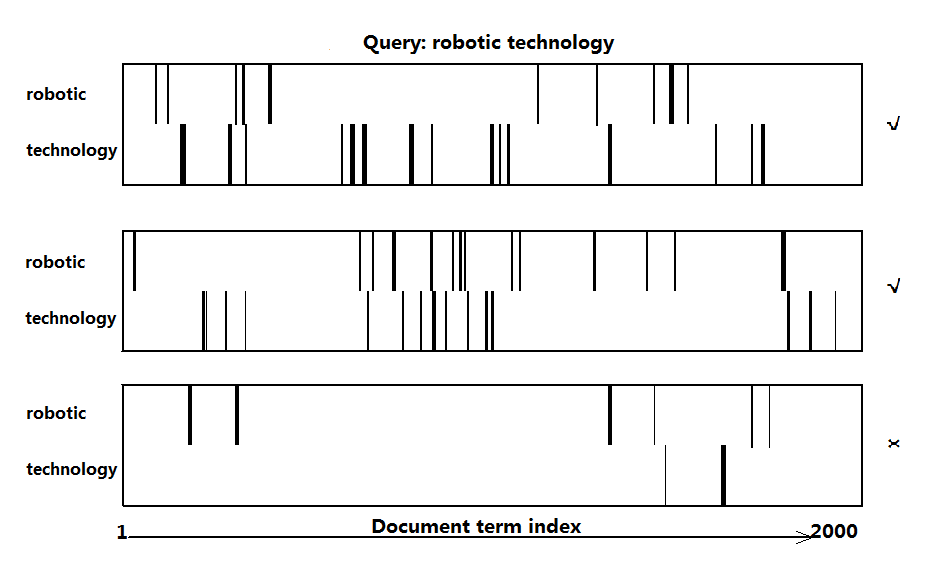}
\caption{Term relevance distribution. The vertical axis denotes the query term, and the horizontal axis denotes the term position index. In each box, the upper part shows the terms related to the query term ``robotic'', and the lower part shows that to ``technology''. The thickness of the line indicates the relevance score of the term, and the thicker the higher. The first two documents are rated relevant by human judges whereas the third one is irrelevant.}\label{fig1}
\end{figure}

To remedy the shortcomings of the previous methods, in this paper, we propose a salient-context-based semantic matching model. With this model, we improve the relevance ranking in IR. Fig.~\ref{fig1} explains the concept of salient context with an example. We have the query terms ``robot technology'' and a corpus of three documents. The three boxes in the figure correspond to those three documents. The vertical lines indicate positions in the documents which are salient with respect to the two query terms, and thus give locations of the salient context.

We can observe that the highly relevant terms are clustered in the first two boxes, while they are scattered in the third. As the two corresponding documents are labeled related to ``robot technology'' by a human, the clustering indicates that the closer together query-related terms are located, the more relevant the document is to the query. This behavior leads us to define the locations of these clusters as the salient context. Our goal is to find the most salient context and embed it into vectors that represent the document. In this way, we eliminate the risk of single-keyword mismatching, thus addressing the shortcomings of the models mentioned earlier.

To locate the most salient context, we define a measurement of the contextual salience. It is based on the semantic similarity between the query and the salient context and is designed such that it is not influenced by low query-related terms or dominated by a single term. In addition, we use the BM25 relevance score as a representation of the global context in the final relevance function.  

This paper has threefold contribution. Firstly, we analyze and demonstrate the aggregation phenomenon of highly query-related terms in relevant documents, and also define our new concept of salient context. Secondly, we propose a way to measure contextual-salience to locate the most salient context dynamically. Thirdly, rather than using the context surrounding a keyword, we propose to use the most salient context as a representation of a document, thereby eliminating the mismatching problem.

\section{Methodology}
\subsection{Term-level Semantic Matching}
\begin{figure}[h]
\centerline{\includegraphics[scale=0.6]{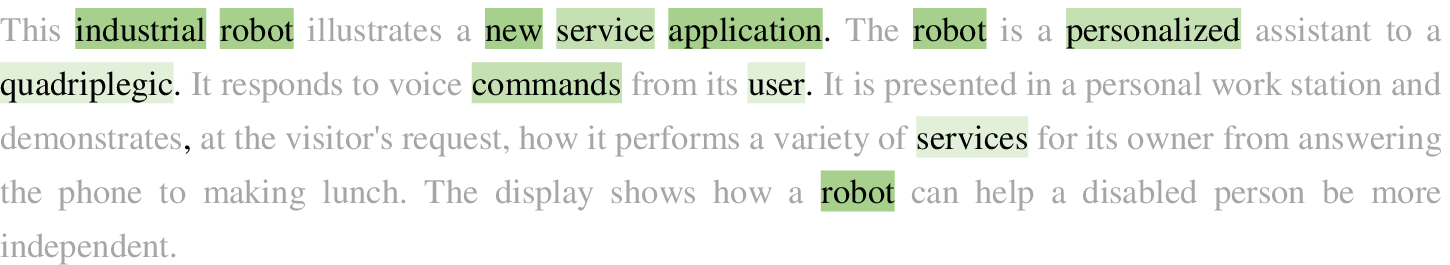}}
\caption{Analysis of term importance for estimating the relevance of a document to the query ``robot technology'' by semantic relevance matching.} \label{fig2}
\end{figure}
Generally, it is important that each keyword is exactly matched. It is often particularly important when the keywords are new or rare. However, traditional keyword matching might lose to capture the fine-grained contextual information and semantically related terms. As illustrated by the example in Fig.~\ref{fig2}, semantic relevance matching is able to highlight the terms with a high semantic relevance to the query ``robotic technology'' with dark green being most relevant. We can see that the semantic matching gives emphasis to semantic related terms such as ``robot'', ``industrial'' and ``application''.

Distributed representations of text, \textit{i.e.} word embeddings, encapsulate useful contextual information and effectively represent the semantic information of a word. Models that use pre-trained word embeddings\cite{DRMM,hui2017pacrr,mcdonald2018deep} have shown better performance than those which use term co-occurrence counting between query and documents. Inspired by this, we utilize the pre-trained word embeddings as the basis for our semantic representation to model the query-document matching interaction. From the embedded vectors, We apply cosine similarity to the capture of the word-level semantic matching as given by:
\begin{equation}
\begin{aligned}
 & s_{ij} = \frac{(\bm{w_i})^T\bm{w_j}}{\|\bm{w_i}\|\cdot\|\bm{w_j}\|}\, ,\\
 \end{aligned}
\end{equation}
 where $\bm{w_i}$ and $\bm{w_j}$ represent the vectors for the i-th query term and the j-th document term, respectively.

\subsection{Contextual Salience} %
According to the query-centric assumption proposed in~\cite{wu2007retrospective}, the local context surrounding the location of a found query term in a document is relevant when deciding if the document is a match to the query. In Fig.~\ref{fig2}, relevant terms cluster around the first two sentences, and in Fig.~\ref{fig1} we can see that these clusters are present at both the beginning, middle, and end of a document. Thus, the position of the salient context changes from document to document and therefore our salience-measure must be able to handle that shift. We use a sliding window which moves over the document from the start to the end. For a given position of the window, terms which are highly related to the query are found and thus that part of the document will stand out. The window context for the i-th query term is described as:

\begin{equation}
\begin{aligned}
& \mathbb{S}_i = \{s_{ij} \lvert \,  i\in Q, j\in T\}\,,\\
\end{aligned}
\end{equation}
where $s_{ij}$ is the cosine distance between the i-th query term and the j-th document term in the window, $Q$ is the set of query terms, $T$ is the set of document terms in the window, and $\mathbb{S}_i$ represents the cosine relevance between the i-th query term and the document terms which falls inside the window.

This approach is different from the deep learning models. As stated above, the deep learning models combine all terms in a document into one single document representation. Our representation only takes the relevant parts of the document and embeds those into a document representation. Often, only a few terms with a high windows relevance score contribute to the final document relevance. In order to filter away text noise and counteract semantic drift, we choose to only take the window contextual salience of the top $n$ semantic relevance matches into account. \cut{ Formally, we define contextual salience as follows:
\begin{definition}
Contextual Salience: Given a query and a document, with a fixed-size window shifting from the beginning to the end of the document, if the sum of the semantic similarity between all query terms and the document terms within a certain window surpasses that of all other windows, then the text within this window is defined as the most salient context in the document. The sum of the semantic similarity between all query terms and the document terms within this window is called contextual salience. 
\end{definition}
The mathematical definition is:} 
Here is the processing for getting the n-maximums of the set $\mathbb{S}_i$.
\begin{align}
    S_i^{(1)} &= max(\mathbb{S}_i)\nonumber \\
    S_i^{(2)} &= max(\mathbb{S}_i \setminus \{S_i^{(1)}\})\nonumber \\
    S_i^{(3)} &= max(\mathbb{S}_i \setminus \{S_i^{(1)}, S_i^{(2)}\})\nonumber \\
    \vdots\nonumber\\
    S_i^{(n)} &= max(\mathbb{S}_i \setminus \{S_i^{(1)}, S_i^{(2)}, \cdots, S_i^{(n-1)}\})\, .
\end{align}
The set of n-maximum members of the set is $\mathbb{S}_i$ then
\begin{align}
    \mathbb{S}_i^{n} &= \{ S_i^{(1)}, S_i^{(2)}, \cdots, S_i^{(n)} \}\, .
\end{align}
\cut{
Another, more convoluted, way of writing almost the same using set-notation:
Initially, we define $\mathbb{S}_i^{(0)}=\mathbb{S}_i$ which allows us to get the first maximum value of $\mathbb{S}_i$ as:
\begin{align}
    \mathbb{M}_i^{(1)} &= \{ m\in\mathbb{S}_i^{(0)} : m \geq s \forall s\in \mathbb{S}_i^{(0)} \}
\end{align}
We then define $\mathbb{S}_i^{(1)}=\mathbb{S}_i^{(0)} \setminus \mathbb{M}_i^{(1)}$, and by induction we write:
\begin{align}
    \mathbb{M}_i^{(n+1)} &= \{ m\in\mathbb{S}_i^{(n)} : m \geq s \forall s\in \mathbb{S}_i^{(n)} \}
\end{align}
with $\mathbb{S}_i^{(n+1)}=\mathbb{S}_i^{(n)} \setminus \mathbb{M}_i^{(n)}$. Then, $\mathbb{M}_i^{(n+1)}$ is the set of the $n+1$ maximum values of the set $\mathbb{S}_i$.
}

\begin{equation}
\begin{aligned}
& S_i^{\text{ct}} =  S_{i}^{(1)} +\alpha \cdot \frac{\sum_{n=1}^{K} S_{i}^{(n)}}{K} \, , \\
\end{aligned}
\end{equation}
where $K$=$log(L)+1$, decided by window width $L$. 
$\alpha$ is the influence factor to balance semantic interactions' weighting in the window context.

Queries used in IR are short and without complex grammatical structures. Consequently, we need to take the term importance into account. The compositional relation between the query terms is usually the simple ``and'' relation when searching. Take the given query ``arrested development'' for example, a relevant document should refer to ``arrested'' and ``development'', where the term ``arrested'' is more important than ``development''. There have been many previous studies on retrieval models showing the importance of term discrimination ~\cite{fang2011diagnostic}.
In the proposed model, we introduce an aggregation weight for each query term which controls how much the relevance score on that query contributes to the final relevance score:
\begin{equation}
\begin{aligned}
 & S^{\text{ct}}= \sum_{\mathclap{i=1}}^{ql} g_i S_i^{\text{ct}} \, ,\\
 \end{aligned}
\end{equation}
\begin{equation}
\begin{aligned}
 & g_i =\frac{exp(\bm{v_i}^\text{T}\bm{w_i})}{\sum_{m=1}^{ql} exp(\bm{v_m}^\text{T}\bm{w_m})}\,  \label{equ:weight} ,
 \end{aligned}
\end{equation}
where $\bm{v}_i$ denotes the weight vector of the i-th query term vector $\bm{w_i}$, and $ql$ is the query length. In our model, we set the weight vectors equal to their respective query term vector, \textit{i.e.} $\bm{v_{i}}= \bm{w_{i}}$. Putting this into Equ.~\ref{equ:weight}, we get:
\begin{align}
    g_i &= \frac{exp(\bm{w}_i^\text{T}\bm{w}_i)}{\sum_{m=1}^{ql} exp(\bm{w}_m^\text{T}\bm{w}_m)}\, \label{equ:final_softmax}
\end{align}
Here, $\bm{w}_i^\text{T}\bm{w}_i$ squares each element of $\bm{w}_i$ before summing them together. As $\bm{w}_i\in [-1,1]^{d}$, with $d$ being the dimension of the weight vector, the resulting scalar will be positive and equal to the square of the magnitude of $\bm{w}_i$. Equ.~\ref{equ:final_softmax} is the normalized exponential, or softmax, function, with $g_{i}\in [0,1]$. It returns a scalar which is proportional to the normalized magnitude of the term vector, but with an emphasis on the vectors with the largest magnitudes. Thus, it regularizes the relevance score.

\subsection{Relevance Aggregation}

Different from semantic-matching-based distributional word embedding, exact keywords matching avoids the risk of rare or new words in query. Hence, we linearly combining the exact keywords matching and use it as a compensation for semantic matching. Traditional IR models ,such as BM25~\cite{robertson2009probabilistic}, is a classical weighting function employed by the Okapi system. As shown by previous TREC experimentation, BM25 usually provides very effective retrieval performance on the TREC collections. In BM25, the relevance score is based on the within-document term frequency and query term frequency. We can utilize BM25 to model relevance matching in document-level with query terms. In our paper, we apply BM25 to extend model on document-level matching and define the way to aggregate exact keywords matching interactions by integrating into BM25 linearly via a parameter $\beta$. We also take into consideration of the co-occurrence of query terms within document in weighting function for the contextual salience in the document. The two formulas are defined as below: 

\begin{equation}
\begin{aligned}
& Linear\, Function: \mathcal{F}(S^{ct}) = max(S^{ct}) + \beta \cdot BM25\,, \\
\end{aligned}
\end{equation}
\begin{equation}
\begin{aligned}
CO\, Weighting\, Function: \mathcal{F}(S^{ct})&=\log(co+C) \cdot \max(S^{ct}) \\
 &\,\,+\beta \cdot BM25\,, C\in \mathbb{R}\,,\\
\end{aligned}
\end{equation}
where $\beta$ is the influence factor to balance BM25, decides the effects of BM25 in relevance scoring. When $\beta$ is 0, only contextual salience contributes the relevance scoring, $\beta \in$ (0,1) the contextual salience and BM25 contribute the relevance scoring together. 
$co$ is the co-occurrence of query terms within document, and the constant $C$ is a constant to balance parameter $co$. 

\section{Data Sets and Evaluation}

We evaluate the proposed approach on five standard TREC collections , which are different in their sizes, contents, and topics. The TREC tasks and topic numbers associated with each collection are summarized in Table~\ref{tab2}.
\begin{table}[htbp]
\centering
\caption{Overview of the TREC collections used}\label{tab2}
\begin{tabular}{|c|c|c|c|}
 \hline
\textbf{Collection Name}& \textbf{Topics}& \textbf{Topics Num.}& \textbf{Docs}\\
 \hline
AP8889  & 51-100 &50 &164,597\\
 \hline
WT2G &  401-450 &50 &247,491\\
 \hline
Robust04 & 301-450 601-700 &250 &528,155\\
 \hline
 WT10G  & 451-550 &100 &1,692,096 \\
 \hline
 Blog06 &851-950 &100 &3,215,171  \\
 \hline
\end{tabular}
\end{table}
For all the test collections used in our experiments, we apply pre-trained GloVe word vectors\footnote[1]{\scriptsize{https://nlp.stanford.edu/projects/GloVe/}} which are trained from a 6 billion token collection (Wikipedia 2014 plus Gigawords 5), reliable term representations can be better acquired from large scale unlabeled text collections rather than from the limited ground truth data for IR task. We use the TREC retrieval evaluation script\footnotetext[2]{\scriptsize{https://trec.nist.gov/trec\_eval/}} focusing on MAP, RP (recall precision) and P@5, P@20, NDCG@5, and NDCG@20 in our experiments.
We provide the source code\footnote[3]{\scriptsize{source code is available on https://github.com/YuanyuanQi/CSSM\_IR/}} for the model as well as trained word vectors.

\section{EXPERIMENTS}
\begin{table*}[htbp]
\caption{Comparisons of CSSM and BM25, with MAP, RP and P@5, P@20, NDCG@5, and NDCG@20 over five TREC collections}\label{tab3}
\centering
\begin{tabular}{|c|c|c|c|c|c|c|c|c|}
\hline
  \textbf{Corpus} &\textbf{Methods} & \textbf{MAP} & \textbf{RP} & \textbf{P@5}& \textbf{P@20} & \textbf{NDCG@5} & \textbf{NDCG@20} \\ 
\hline
 AP8889 & BM25 & 0.278  & 0.298  & 0.453  & 0.404  & 0.461  & 0.430 \\
& $CSSM_{lf}$  & 0.298  & 0.319  & 0.490  & 0.413  & 0.494  & 0.447 \\
&\, & +\textbf{7.26}\% & +\textbf{7.04}\% & +\textbf{8.10}\% & +\textbf{2.28}\% & +\textbf{7.15}\% & +\textbf{3.91}\% \\
& $CSMM_{cw}$ & 0.298 & 0.318  & 0.490  & 0.414  & 0.497  & 0.449 \\
&\, & +\textbf{7.12}\% & +\textbf{6.44}\% & +\textbf{8.10}\% & +\textbf{2.52}\% & +\textbf{7.72}\% & +\textbf{4.30}\% \\
 \hline
WT2G & BM25 & 0.313  & 0.340  & 0.532  & 0.391  & 0.542  & 0.470 \\ 
& $CSSM_{lf}$  & 0.366  & 0.373  & 0.596  & 0.421  & 0.611  & 0.514  \\ 
&\, & +\textbf{17.05}\% & +\textbf{9.83}\% & +\textbf{12.03}\% & +\textbf{7.67}\% & +\textbf{12.69}\% & +\textbf{9.48}\% \\ 
& $CSMM_{cw}$ & 0.368  & 0.378  & 0.600  & 0.428  & 0.616  & 0.521 \\  
&\, & +\textbf{17.82}\% & +\textbf{11.09}\% & +\textbf{12.78}\% & +\textbf{9.46}\% & +\textbf{13.63}\% & +\textbf{10.88}\% \\ 
 \hline
Robust & BM25 & 0.239  & 0.283  & 0.481  & 0.354  & 0.497  & 0.425 \\  
& $CSSM_{lf}$  & 0.261  & 0.304  & 0.497  & 0.374  & 0.508  & 0.443 \\  
&\, & +\textbf{9.60}\% & +\textbf{7.50}\% & +\textbf{3.35}\% & +\textbf{5.68}\% & +\textbf{2.11}\% & +\textbf{4.24}\% \\ 
& $CSMM_{cw}$ & 0.262  & 0.304  & 0.496  & 0.376  & 0.508  & 0.445 \\  
&\, & +\textbf{9.94}\% & +\textbf{7.72}\% & +\textbf{3.18}\% & +\textbf{6.19}\% & +\textbf{2.13}\% & +\textbf{4.69}\% \\ 
 \hline
 WT10G & BM25 & 0.211  & 0.244  & 0.382  & 0.274  & 0.418  & 0.362 \\  
& $CSSM_{lf}$ & 0.224  & 0.257  & 0.394  & 0.282  & 0.432  & 0.373 \\  
&\, & +\textbf{6.25}\% & +\textbf{5.34}\% & +\textbf{3.22}\% & +\textbf{3.14}\% & +\textbf{3.57}\% & +\textbf{3.01}\% \\ 
& $CSSM_{cw}$ & 0.221  & 0.255  & 0.404  & 0.276  & 0.447  & 0.371 \\  
&\, & +\textbf{4.55}\% & +\textbf{4.64}\% & +\textbf{5.90}\% & +\textbf{0.73}\% & +\textbf{6.99}\% & +\textbf{2.40}\% \\ 
 \hline
 Blog06 & BM25 & 0.318  & 0.371  & 0.634  & 0.605  & 0.625  & 0.611 \\  
& $CSSM_{lf}$ & 0.345  & 0.403  & 0.664  & 0.642  & 0.653  & 0.647 \\  
&\, & +\textbf{8.53}\% & +\textbf{8.69}\% &+\textbf{ 4.73}\% & +\textbf{6.12}\% & +\textbf{4.53}\% & +\textbf{5.96}\% \\ 
& $CSMM_{cw}$ & 0.346  & 0.403  & 0.670  & 0.642  & 0.659  & 0.648 \\  
&\, & +\textbf{8.748}\% & +\textbf{8.79}\% & +\textbf{5.68}\% & +\textbf{6.03}\% & +\textbf{5.48}\% & +\textbf{6.09}\% \\
 \hline
\end{tabular}
\end{table*}

Table~\ref{tab3} shows the performance comparisons between the baseline model BM25 and new model CSSM on five collections over MAP, RP and P@5, P@20, NDCG@5 and NDCG@20. The percentage of how much our model outperforms BM25 is also listed. With regards to MAP and RP it indicates that, in general our model performs better than the baseline model BM25 on all five collections, especially on WT2G, Robust04 and Blog06 collections. It demonstrates the importance of semantic relevance matching and emphasizes contextual salience is helpful to locate the most relevant local context through highly semantic relevance matching.  Compare the results of $CSSM_{lf}$ (linear function) and $CSSM_{cw}$(co weighting function), three datasets show improvements, the co-occurrence information of query terms in document can offer positive connection with contextual salience in the model. The experiment results prove that our model can encode the critical contextual semantic information in our relevance ranking function for the IR.

\begin{table}[htbp]
\centering
\caption{Comparisons of Deep Learning methods on Robust04 collection}\label{tab4}
\begin{tabular}{|c|c|c|c|}

\hline
  \textbf{Methods} & \textbf{MAP} & \textbf{P@20} & \textbf{NDCG@20} \\ 
  \hline
  BM25 & 0.239  & 0.354  & 0.425 \\ 
\hline
  DRMM & 0.256  & 0.37  & 0.444 \\ 
 \hline
  PACRR & 0.258 & 0.372  & 0.443 \\ 
 \hline
 DRMM+PACRR & 0.259 &  0.372 & 0.444 \\ 
 \hline
 ABEL-DRMM & 0.263 & 0.380 & 0.456 \\ 
 \hline
ABEL-DRMM+MV & 0.265 & 0.380 & 0.455 \\ 
\hline
  $CSSM_{lf}$ & 0.266  & 0.379  & 0.455\\
 \hline
  $CSSM_{cw}$ & 0.266  & 0.380  & 0.455\\
 \hline
POSIT-DRMM  & 0.270 & 0.383 & 0.457\\
\hline
POSIT-DRMM+MV  & 0.272 & 0.386  & 0.461\\
 \hline
\end{tabular}
\end{table}

Table~\ref{tab4} shows the performance on Robust04 collection with comparison of deep learning based methods recently proposed in\cite{DRMM,hui2017pacrr,mcdonald2018deep}. 
Our performance is better than DRMM, PACRR, DRMM-PACRR, slightly better than ABEL-DRMM and ABEL-DRMM+MV with less extra model training data.  Compare with POSIT-DRMM and POSIT-DRMM+MV which encode multiple views (MV) of terms (context-sensitive term encodings, pre-trained term embeddings, and one-hot term encodings), our model utilizes pre-trained term embeddings alone. We mainly take into account of two reasons. First, according to our scoring function, directly applying multiple views of terms is hard to balance the input dimensions differences, one-hot vector is high dimensional and sparse term embedding. Second, it needs sacrifice efficient to take training data to explicitly tune context-sensitive term encodings in model. In addition, without model parameters tuning, our model retrieval time costing is less than all supervised deep learning based models in the table, works as efficiently as BM25.

\section{CONCLUSION}
In this paper, we propose a semantic-matching based method to locate the most salient context for understanding a piece of multimedia content. We propose to prioritize the action of locating the semantic salient context in the relevance calculation. On the basis of the prioritization, we define a measurement of contextual salience to quantify the relevance of a document towards a query. Furthermore, we apply the proposed method in IR, and it shows promising improvements over the strong BM25 baseline and several neural relevance matching models. Finally, extensive comparisons between several neural relevance matching models and our approach suggest that explicitly modelling the salient query-related context in document is helpful to improve the effectiveness of relevance ranking for IR. Our idea of understanding content by locating the most salient context provides a new perspective in multimedia content analysis, and the proposed semantic-matching based method can be applied to other forms of multimedia content. The proposed method provides an efficient and explainable relevance ranking solution which can be generalized to other forms of multimedia content as well.

\section{ACKNOWLEDGE}
This work was supported by Beijing Natural Science Foundation (4174098), National Natural Science Foundation of China (61702047), National Natural Science Foundation of China (61703234) and the Fundamental Research Funds for the Central Universities (2017RC02).

\vspace{12pt}
\color{red}

\end{document}